\documentclass[manuscript]{aastex631}   

\usepackage{booktabs}  
\usepackage{tabularx}
\usepackage{amsmath}  
\usepackage{multirow} 
\usepackage{turnstile} 
\usepackage{graphicx}  
\usepackage{array}
\usepackage{hyperref}


\defcitealias{FrewDavid2016MNRAS}{FW16}
\defcitealias{Ali2015ApSS}{A15} 
\defcitealias{AliA2022RAA}{A22}
\defcitealias{ZhangCY1995ApJ}{Z95}
\defcitealias{CahnKalerStanghellini1992AAS}{CKS92}
\defcitealias{vdsz1995AA}{Vdsz95}
\defcitealias{caGaiaDR22018AJ}{CGDR2}
\defcitealias{caGaiaEDR32021AJ}{CGDR3}
\shorttitle{FAST search for circumstelar \ion{H}{1}: Bubbles}
\shortauthors{Liu et al.}
\begin{document}

\title{FAST search for circumstellar atomic hydrogen. IV.
 bubbles associated with planetary nebulae}

\correspondingauthor{Yong Zhang}
\email{zhangyong5@mail.sysu.edu.cn}

\author[0009-0007-1038-4254]{Jun-Yang Liu}
\affiliation{School of Physics and Astronomy, Sun Yat-sen University, 2 Daxue Road, Tangjia, Zhuhai, Guangdong Province,  China}

\author[0000-0002-1086-7922]{Yong Zhang}
\affiliation{School of Physics and Astronomy, Sun Yat-sen University, 2 Daxue Road, Tangjia, Zhuhai, Guangdong Province,  China}
\affiliation{CSST Science Center for the Guangdong-Hongkong-Macau Greater Bay Area, Sun Yat-Sen University, Guangdong Province, China}

\author[0000-0002-2762-6519]{Xu-Jia Ouyang}
\affiliation{College of Physics, Guizhou University, Guiyang, Guizhou Province, China}




\begin{abstract}
Investigating the bubbles generated by the interaction between asymptotic giant branch  stellar outflows and the interstellar medium (ISM) is pivotal for elucidating the mechanism by which evolved low- to intermediate-mass stars enrich the ISM with heavy elements. 
Using archival datasets from the Galactic Plane Pulsar Snapshot survey and the Galactic Arecibo L-Band Array  \ion{H}{1} survey, we have identified 14  bubbles within interstellar atomic hydrogen (\ion{H}{1}) maps,
each showing evidence of potential association with planetary nebulae (PNe).
We pursue two primary objectives centered on the identified ISM bubbles and their association with PNe. First, leveraging the calibrated distance measurements of PNe from Gaia Data Release 3, we utilize these ISM bubbles as observational tracers to investigate and constrain the Galactic rotation curve. 
Second, we note that distance determinations for some PNe remain unreliable, partly because their central stars are obscured by extended nebular envelopes
or are misidentified. Therefore, we develop a novel methodological framework to derive kinematic distances for PNe by leveraging the velocities of their associated ISM bubbles and constraints from the Galactic rotation curve.
\end{abstract}


\section{Introduction} \label{sec:intro} 
Galactic interstellar bubbles are ubiquitous dynamically significant structures throughout the Milky Way that serve as key astrophysical probes to unravel the physical processes of stellar feedback and the cyclic exchange of matter and energy between evolved stars and the interstellar medium (ISM). These bubbles act as ``tracer fossils'' of past stellar activity because they record how mass ejected by stars either steadily via winds or impulsively via explosive events interacts with the ambient ISM, shaping the structure of the interstellar environment and regulating the conditions for subsequent star formation. Such bubbles originate from  energetic sources including the fast dense winds of massive O- and B-type stars, the slow dusty winds of asymptotic giant branch (AGB) stars which are low- to intermediate-mass stars in their final evolutionary stages, and the powerful shockwaves driven by supernova explosions. Among these the link between AGB stars and planetary nebulae (PNe) is particularly intimate as PNe themselves are the end products of AGB evolution. 
The interacting stellar winds (ISW) model, which serves as a cornerstone of PN formation theory, has been widely accepted and empirically validated over the past several decades \citep{Kwoks1982ApJ,Balick1987AJ}.
This model describes a two-stage process in which during the late AGB phase the star ejects a slow, dense wind with a velocity of $\sim10$ km s$^{-1}$ and as the star evolves into a hot central star of a PN it emits a fast, tenuous wind at a velocity of $\sim10^3$ km s$^{-1}$.
The collision between these two winds generates a strong shock front that sweeps up the surrounding material carving out a circumstellar bubble that eventually evolves into the ionized morphologically diverse shell of a PN. Consequently PNe are
not merely endpoints of stellar evolution but also invaluable probes for studying circumstellar bubbles sculpted by stellar winds.
In the present study, we focus on a less explored yet critical aspect of this evolutionary chain, namely the long-term interaction between AGB stellar winds and the ambient ISM both prior to and during the PN phase. 
This interaction does not vanish when the star transitions to the PN stage; instead, it imprints distinct, measurable signatures in both the spatial distribution and kinematic properties of interstellar atomic hydrogen (\ion{H}{1}) emission. These \ion{H}{1} signatures persist into the PN phase, thereby establishing a unique observational link between the AGB wind era and the mature PN stage, one that we leverage to address key questions regarding Galactic dynamics and PN distance measurements.

The concept of interactions between PNe and the ISM was  put forward by \cite{Gurzadian1969NewYork}, laying the initial theoretical foundation for this research field; building on this early proposal, \cite{SmithH1976MNRAS} developed the first quantitative framework by adopting a thin-shell approximation for PNe and integrating it with the momentum-conserving ``snowplow'' model adapted from \cite{Oort1951pca}’s theory, a step that transitioned the topic from qualitative speculation to a structured, mathematically describable research direction. Following the establishment of this quantitative basis, subsequent progress focused on refining physical understanding through hydrodynamic simulations: \cite{Borkowski1990ApJ} and \cite{Soker1991AJS} extended earlier concepts by simulating PN-ISM interactions across the full evolutionary sequence from the AGB phase to the mature PN phase, and a critical finding from their work was that such interactions only become observable when nebular density drops below a critical threshold, a condition generally satisfied in late-stage PNe or extended Galactic disk nebulae, which clarified the specific astrophysical scenarios where these interactions are relevant and narrowed the scope for subsequent observational studies. Complementing these theoretical and simulation-based advancements, \cite{Zijlstra2002ApJ} later provided observational evidence for PN-ISM interactions, validating the phenomenon proposed in earlier models; further refinements to dynamic understanding followed as \cite{Villaver2003ApJ}, via two-dimensional numerical simulations, demonstrated that PN-ISM interactions are not isolated events but occur continuously throughout the AGB-to-PN evolutionary process, updating the earlier view of interactions as ``late-stage only'' and emphasizing their persistent role in shaping nebular evolution. The theoretical framework was then further expanded by \cite{Wareing2007MNRAS}, who employed three-dimensional hydrodynamical simulations and introduced a ``triple-wind'' model,  encompassing AGB superwinds, post-AGB winds, and ISM flows, to more comprehensively capture the multi-component dynamics of PN-ISM interactions, and notably, interstellar \ion{H}{1} serves as a critical diagnostic tool in this context: as a multi-scalar tracer of the ISM \citep{Kalberla2009ARAA}, it enables researchers to probe both large-scale Galactic dynamics and localized stellar feedback processes, bridging the gap between theoretical simulations of PN-ISM interactions and observational constraints.

The 21 cm line, endowed with the unique capability to probe both large-scale Galactic dynamics and sub-parsec-scale interfaces between stellar winds and the ISM, stands as a powerful tool for the systematic identification of stellar wind cavities. Single-dish observations conducted by \cite{ouyang2023ApJ} demonstrate that the 21 cm emission line of interstellar \ion{H}{1} is capable of tracing interactions between stellar winds and the ISM.
When applied to bubbles associated with PNe, such \ion{H}{1} observations enable the derivation of kinematic distances to PNe that are independent of assumptions inherent to nebular models or empirical scaling relations effectively addressing key limitations of traditional statistical distance methods. In turn, the bubbles sculpted by PN–ISM interactions serve as independent tracers of Galactic kinematics, presenting new opportunities to test the validity of existing Galactic rotation models (GRMs) and refine their parameters for greater precision.

The GRM is a fundamental tool for probing the Milky Way’s mass distribution and dynamics; traditionally, in the solar neighborhood, the profile of GRM and Galactic parameters have always been determined by the tangent-point (TP) approach, which is based on the radio observations of \ion{H}{1} at 21\,cm and CO at 2.6\,mm gas emission \citep[e.g.,][]{Burton1978AAGRM, Clemens1985ApJGRM, Fich1989ApJGRM, Levine2008ApJ, Sofue2009PASJGRM}. The TP method, however, relies on the assumption of purely circular orbits, and its application becomes unreliable in the innermost Galactic region ($R \leq 4$–$5\,\rm kpc$), where gas kinematics are strongly perturbed by the Galactic bar. For the outer region ($\rm R \gg R_0$), the GRM is determined by different tracers associated with kinematic properties and known distances, such as red clump giants \citep{Bovy2012ApJ,Lopez2014A&A, Huang2016MNRAS}, \ion{H}{2} regions \citep{Fich1989ApJGRM, BrandGRM1993AA}, classical Cepheids (as standard candles) \citep{Pont1997A&A}, PNe \citep{Pont1997A&A}, masers \citep{Reild2009ApJGRM, Honma2012PASJ, 2014GRMReid2014ARAA, ReildGRM2019ApJ}, blue horizontal branch stars \citep{Xue2008ApJ, Deason2012MNRAS, Kafle2012ApJ}, and globular clusters \citep{Spitler2009MNRAS, watkins2010MNRAS, Li2017ApJ,Li2020ApJ}. Among these tracers, significant progress has been achieved with Very Long Baseline Interferometry (VLBI) observations of masers in star-forming regions, which provide precise distances and proper motions and thus yield a more accurate determination of the Galactic constants and the rotation curve. Even these state-of-the-art measurements, however, are limited by sample size and sky coverage, so it is crucial to explore complementary tracers.

Obtaining reliable distances to Galactic PNe remains challenging due to the wide diversity of their central stars: individual methods with high/moderate accuracy include trigonometric parallax \citep[for nearby PNe; see, e.g.,][]{Harris2007AJ,Benedict2009AJ,Harrison2013ApJ}, expansion parallax \citep[e.g.,][]{schnberner2018A&A}, interstellar reddening analysis \citep[e.g.,][]{Gathier1986AA}, and \ion{H}{1} absorption line measurements \citep[e.g.,][]{Yangay2016ApJs}, with selection contingent on object-specific observables and instrumental constraints; statistical methods based on empirical scaling relations and used for most Galactic PNe have a framework established by \cite{Shklovsky1956AZ}, refined by \cite{Dual1982A&AS} and \cite[][hereafter \citetalias{CahnKalerStanghellini1992AAS}]{CahnKalerStanghellini1992AAS}, and later extended via relations like ionized mass–optical depth \citep{Stanghellini2008ApJ}, 5 GHz radio surface brightness–radius correlations, and ionized mass–radius scaling laws \citep[e.g.,][hereafter \citetalias{vdsz1995AA}, \citetalias{ZhangCY1995ApJ}, \citetalias{Ali2015ApSS}, and \citetalias{AliA2022RAA}, respectively]{vdsz1995AA, ZhangCY1995ApJ, SchneiderBuckley1996ApJ,BensbyTLundstrI2001A&A, Ali2015ApSS,AliA2022RAA}, while \citet[][hereafter \citetalias{FrewDavid2016MNRAS}]{FrewDavid2016MNRAS} further developed an optical statistical method (using H\(\alpha\) surface brightness–radius relations) for 1100 Galactic PNe. Despite their wide use, statistical techniques suffer from systematic biases (oversimplified geometry, limited calibration samples, model dependencies), making it essential to explore multiple independent tracers.

This paper is structured as follows: Section~\ref{sec:obse} details the observational datasets employed in this study, and Section~\ref{sec:iden} outlines the methodology for identifying bubbles. Section~\ref{sec:res} presents the corresponding results, while Section~\ref{sec:apply} explores applications of these bubbles, specifically in assessing the reliability of GRMs and determining the distances of PNe. Finally, Section~\ref{sec:concl} summarizes our key conclusions.

\section{Data and Reduction} \label{sec:obse}

The analysis incorporates 21\,cm \ion{H}{1} data cubes from the Galactic Plane Pulsar Snapshot \citep[GPPS,][]{GPPSsurvey2021RAA} \ion{H}{1} survey. The GPPS survey, conducted using the Five-hundred-meter Aperture Spherical Telescope \citep[FAST;][]{Nan2011IJMPD} L-band 19-beam receiver \citep{FASTLbandreceiver2020RAA}, represents the most sensitive single-dish \ion{H}{1} survey to date. Our selected data subset spans Galactic longitude $33{\degr}\le l \le55{\degr}$ and latitude $-2{\degr}\le b \le2{\degr}$, with an angular resolution of $\sim2{\arcmin}.9$.
The data cubes are characterized by Galactic coordinate grids with a spatial resolution of $1\arcmin \times 1\arcmin$, though the pixel sampling is non-uniform due to the $\sim2\arcmin.9$ beam. Additionally, they exhibit a spectral resolution of 0.1\,km\,s$^{-1}$, covering a velocity range of $-300$\,km\,s$^{-1} \leq V_{\rm LSR} \leq 300$\,km\,s$^{-1}$ in the local standard of rest (LSR) frame.
The RMS brightness temperature noise reaches 40\,mK per channel after standard ON-OFF calibration.

Complementary data comes from the Galactic Arecibo L-Band Feed Array (GALFA) \ion{H}{1} survey \citep{2011ApJS..194...20P,2018ApJS..234....2P}.
Covering $-1{\degr}20{\arcmin}<$ Dec $<38{\degr}02{\arcmin}$ and a full range of RA with a resolution of $4{\arcmin}.9$, its wide database (0.754\,km\,s$^{-1}$ velocity resolution) 
suits our analysis, considering that the typical PN expansion velocities far exceed 1\,km\,s$^{-1}$. 
The GALFA-\ion{H}{1} survey data archive data is given in bright temperature ($T_{\rm B}$).
Flux density conversion is performed using the relation $S_\nu = \int{B_\nu d \Omega}$, where $B_\nu$ is approximated via the Rayleigh-Jeans law as  $B_\nu = {2\nu^2kT_{\rm B}}/{c^2}$ \citep[see][]{ouyang2022ApJ,ouyang2023ApJ}.

\section{Search for \ion{H}{1} bubbles} \label{sec:iden}

\subsection{Sample selection}
We have selected PNe within the GPPS and GALFA \ion{H}{1}
survey regions from the Hong Kong/AAO/Strasbourg H$\alpha$ (HASH) database \citep{hashdatabaseparker2016JPh,hashdatabaseboji2017IAUS}. 
The sample includes sources classified as `true,' `likely,' or `possible' PNe in the HASH catalog. 
Of these, 126 HASH-identified PNe are spatially coincident with the GPPS \ion{H}{1} survey coverage area. 

With a limited angular resolution ($\sim3\arcmin$), single-dish observations primarily sample large-scale bubbles. Can wind–ISM interactions produce cavities of this scale? Such scales are plausible: \citet{IRC10216MNRAS2015} reported a 1300$\arcsec$ (0.8 pc) \ion{H}{1} shell around the AGB star IRC$+$10216, and the Galaxy Evolution Explorer (GALEX) revealed extended PN tails and shells \citep{Sahai2023AJ}. Hydrodynamic simulations further demonstrate that wind–ISM interactions can form bubbles with sizes ranging from 0.1 to 10 pc \citep{wareing2006MNRAs,Wareing2007ApJ}. Collectively, these observational and theoretical results support our identification of the bubbles detected with the FAST as products of PN–ISM interactions. We thus establish the following selection criterion: $r_{\rm exp} \equiv V_{\rm exp}t_{\rm typ}/d > \theta_{\rm beam}$, where $V_{\rm exp} = 20\,\mathrm{km\,s^{-1}}$ denotes the characteristic nebular expansion velocity and $t_{\rm typ} \sim 10^{5}\,\mathrm{yr}$ corresponds to the dynamical timescale of AGB winds. This criterion identifies 11 PNe with well-defined circumstellar bubbles: Sh 2-71, Abell 53, K 3-33, K 3-29, KLSS 1-1, K 3-38, PM 1-288, Pa 131, PM 1-301, PM 1-296, and K 4-28. In addition, three supplemental PNe with apparent cavities (K 3-35, Abell 21, SH 2-68) are located beyond the coverage of the GPPS \ion{H}{1} survey.

\subsection{Identification of ISM bubbles linked to PNe} 

We aim to search for bubbles sculpted by the interaction between AGB/post-AGB stellar winds and the ISM. The identification and characterization of these bubbles can be achieved using  moment-zero maps, channel maps, and position-velocity ($p$-$v$) diagrams. 
For moment-zero maps, we extract $30\arcmin \times 30\arcmin$ sample regions from the data cubes of the GPPS and GALFA \ion{H}{1} datasets, where each region is centered on a selected target; within these maps, we focus on identifying regions of lower \ion{H}{1} column density around PNe relative to the surrounding ISM, as this reduced density serves as a potential indicator of bubble presence. Notably, studies of PN-ISM interactions have demonstrated that cavities may be observed on the downstream side of the PN's proper motion direction \citep[e.g.,][]{Wareing2007MNRAS,Sahai2023AJ}, prompting us to construct $p$-$v$ diagrams along the proper motion vector of each PN, with these diagrams covering an angular scale of $\sim10\arcmin$. By integrating moment-zero maps with $p$-$v$ diagrams, the comprehensive morphology of bubbles encompassing their spatial distribution and kinematic properties can be clearly resolved.

To characterize the reliability of bubble identification, we employ a three-tier confidence scheme (Classes A, B, and C) based on morphological consistency and spatial correlation with PN centroids (Table \ref{Tab: [table:alldistance]}). Class A represents the most robust identifications, defined by a well-defined low-density cavity in moment-zero maps that spatially coincides with the PN, alongside a continuous arcuate structure in $p$-$v$ diagrams within the LSR velocity range and coherent features persisting across multiple consecutive velocity channels in channel maps. Class B candidates meet the core criteria of Class A, including cavity detection in moment-zero maps and consistent $p$-$v$ diagram morphology, but exhibit reduced confidence due to morphological fragmentation in channel maps. Class C, by contrast, designates tentative identifications where bubble-like features appear in either moment-zero maps or $p$-$v$ diagrams, yet lack conclusive evidence of velocity coherence or spatial correlation with the PN; this category specifically highlights potential false positives that require follow-up verification.

After identifying circumstellar bubbles, we characterize their morphological properties via elliptical fitting analysis. This approach is grounded in the spherically symmetric expanding \ion{H}{1} shell model proposed by \citet{Cazzolato2005AJ}, which predicts the formation of a ring-like structure propagating outward from the central region.
Within this framework, the AGB stellar wind emanating from the bubble’s kinematic center interacts with the ISM, imprinting dynamical signatures that are observable through the central velocity component. 
The analysis proceeds by applying elliptical fitting to $p$-$v$ diagrams, in which we identify regions where the \ion{H}{1} column density falls below thresholds corresponding to 4.5\%–9.1\% of the total area-normalized column density distribution.
The uncertainty associated with the derived central velocity, as reported in Table \ref{Tab: [table:alldistance]}, is determined by projecting the major and minor axes of the fitted ellipse onto the velocity dimension, a procedure that quantifies the kinematic ambiguities intrinsic to the bubble’s expansion morphology.


\section{Results}\label{sec:res}

Table~\ref{Tab: [table:alldistance]} summarizes the measured bubble properties. 
Columns (1)--(6)  list PN~G identifier (from the HASH database), source name, Galactic longitude ($l$) and latitude ($b$), cavity central velocity with uncertainty (V$_{\rm c}$), and morphological classification. 
Columns (7)--(15) present kinematic distance determinations, including best-estimate values, uncertainties, prior measurements from the calibrated Gaia DR2/DR3 parallax distances \citep[][hereafter \citetalias{caGaiaDR22018AJ}, \citetalias{caGaiaEDR32021AJ}]{caGaiaDR22018AJ,caGaiaEDR32021AJ}, and statistical distances derived from scale relations. 
To showcase the the procedures involved in our distance measurement approach, the following subsections will offer in-depth analyses of four representative sources.

\subsection{K 3-35}
\begin{figure}[ht!]
\plotone{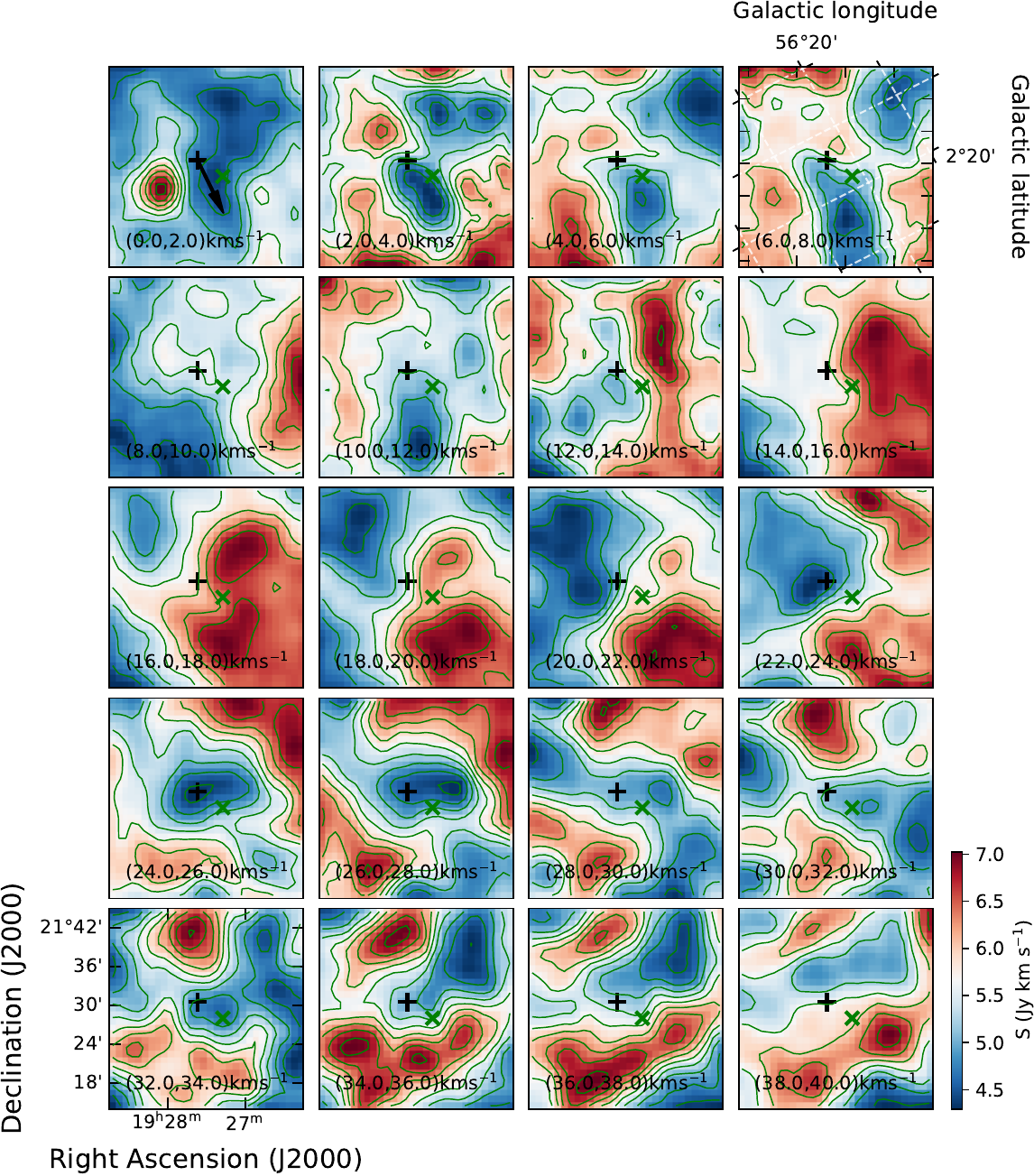}
\caption{Channel maps 
in the velocity range from 0 to 40\,km\,s$^{-1}$ with a channel width of 2\,km\,s$^{-1}$. 
The plus represents the position of K\,3-35. The cross marks the position of a carbon star vul\,10. The galactic coordinate system is shown by the white dashed line in the upper right panel, and the arrow in the upper left panel indicates the proper-motion direction.
\label{fig:vul 10cmap}}
\end{figure}

\begin{figure}[ht!]
\plotone{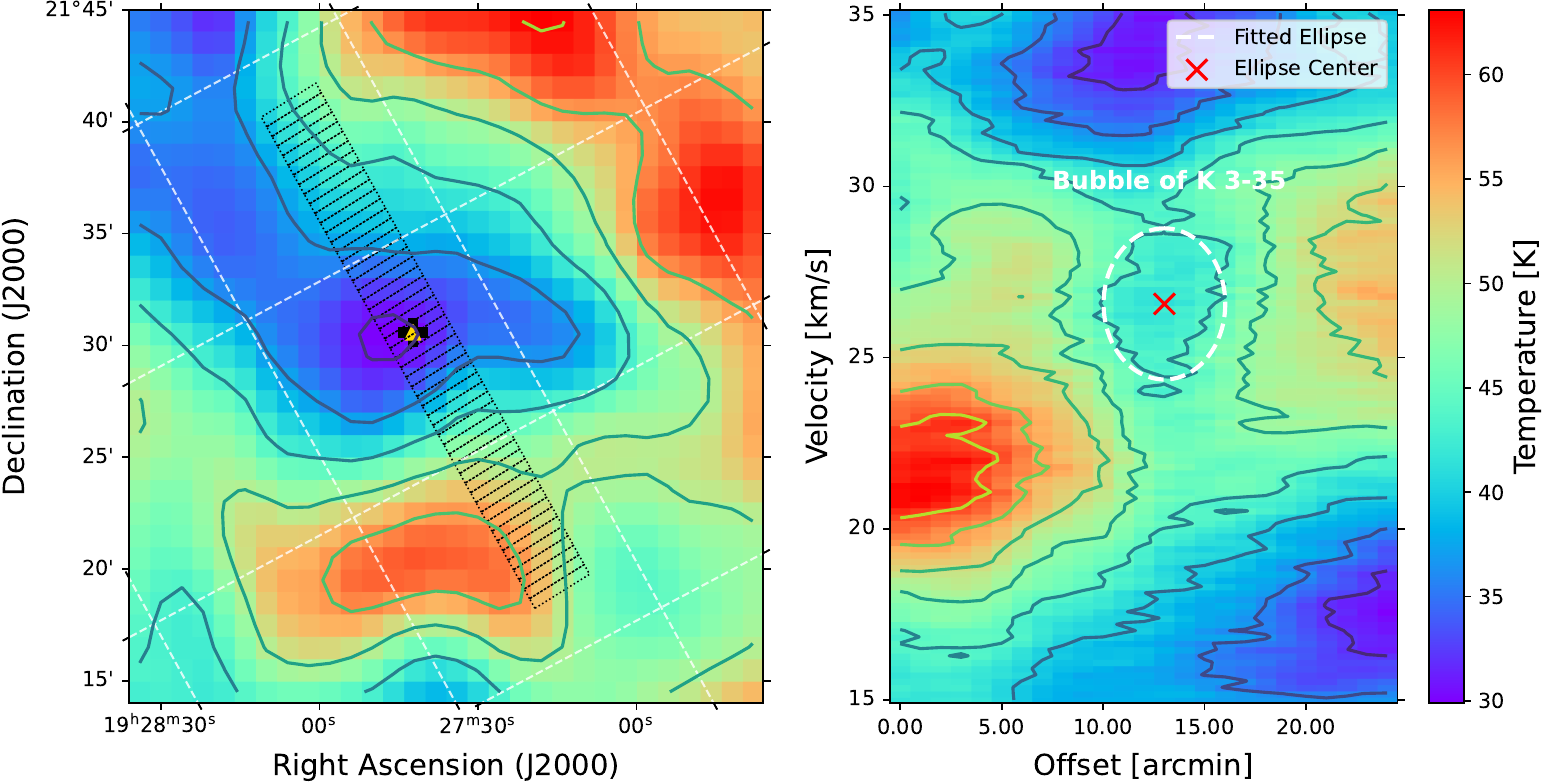}
\caption{Moment-zero map (left) and $\mathit{p\!-\!v}$ diagram (right) of  K 3-35. 
The cross and triangle mark the positions of the bubble
center and PN, respectively.
The approximate extent of \ion{H}{1} absorption associated with the K\,3-35 bubble in the $\mathit{p\!-\!v}$ diagram is shown as a dashed rectangle.
\label{fig:K3-35m0}}
\end{figure}

K\,3-35 displays bipolar outflow morphology in optical imaging \citep{MirandaHbeita2000MNRAS}, while radio continuum observations reveal S-shaped lobe structures with a compact bright core \citep{Miranda2001Nature}. 
The channel maps presented in Figure~\ref{fig:vul 10cmap} delineate the velocity-resolved \ion{H}{1} emission structure across the target field. 
We detect a low \ion{H}{1} column density cavity (Figure~\ref{fig:K3-35m0}, left panel) that fully encompasses the nebula, morphologically consistent with formation via PN-ISM interaction processes. 
The bubble's systemic velocity is unambiguously determined from the $p\!-\!v$ diagram constructed along the nebula's proper motion direction (Figure~\ref{fig:K3-35m0}, right panel). The nearest carbon star Vul\,10 is located at the projected separation of $\sim\,255.7\arcsec$ from K\,3-35 on the sky plane \citep{EDR32021AA}, in Figure~\ref{fig:vul 10cmap}. 
Its systemic velocity of $48.1$ km\,s$^{-1}$ \citep{Schneider1983ApJS} shows a significant discrepancy compared to the nebular velocity of $26$ km\,s$^{-1}$ \citep{2007AJ....133..364T}, with no associated \ion{H}{1} absorption features detected in the moment-0 map. 
These observational constraints conclusively rule out any measurable influence of Vul\,10 on the bubble's properties.

The optimal elliptical fit yields half-maximum radii of $5\arcmin.3$ and $3\arcmin.5$ along the semi-major and semi-minor axes, respectively. 
These values correspond to $9.2\times10^{5}$\,AU and $6.1\times10^{5}$\,AU. 
Observational studies indicate that H$_2$ outflows from AGB stars can extend up to $10^{5}$\,AU \citep{Sahai2023AJ, 2023MNRAS.522..811O, 2023arXiv231009056R}. 
Adopting an expansion velocity of 25\,km\,s$^{-1}$ \citep{MirandaHbeita2000MNRAS}, we calculate the dynamical age of the K\,3-35 cavity as $\tau_{\rm dyn} \sim 10^{5}$\,yr, consistent with typical AGB evolutionary timescales \citep{2018A&ARtypicaltimescale}. 
Hydrodynamic simulations by \citet{Villaver2002ApJ} demonstrate that ISM interactions can decelerate and compress AGB winds, forming extended atomic gas envelopes around molecular components. 
These models predict structure sizes up to $5\times10^{5}$\,AU, in remarkable agreement with our measured \ion{H}{1} cavity dimensions.

\subsection{Sh 2-71}
\begin{figure}[ht!]
\plotone{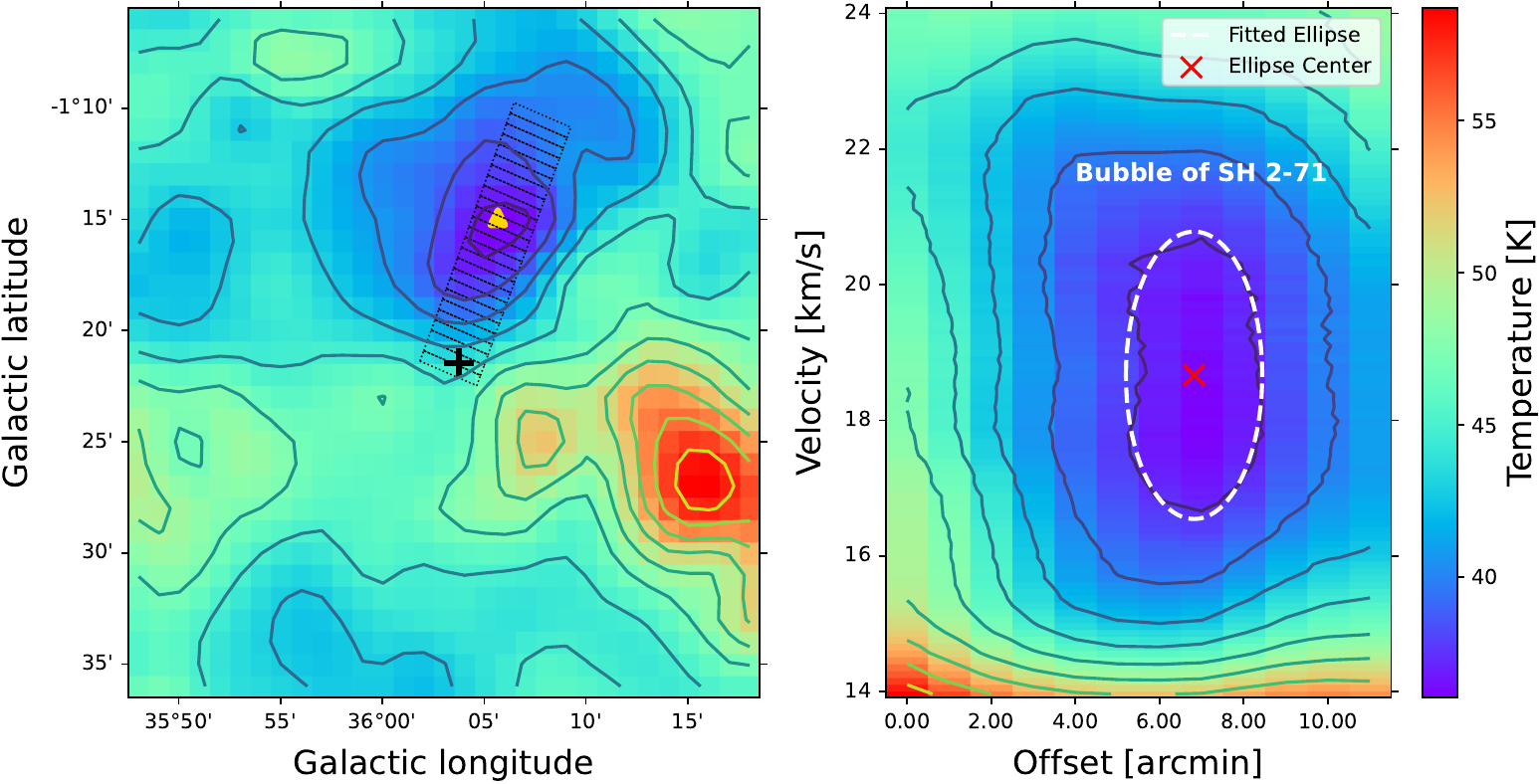}
\caption{Same as Figure~\ref{fig:K3-35m0} but for
Sh 2-71.
\label{fig:SH2-71m0}}
\end{figure}

\begin{figure}[ht!]
\plotone{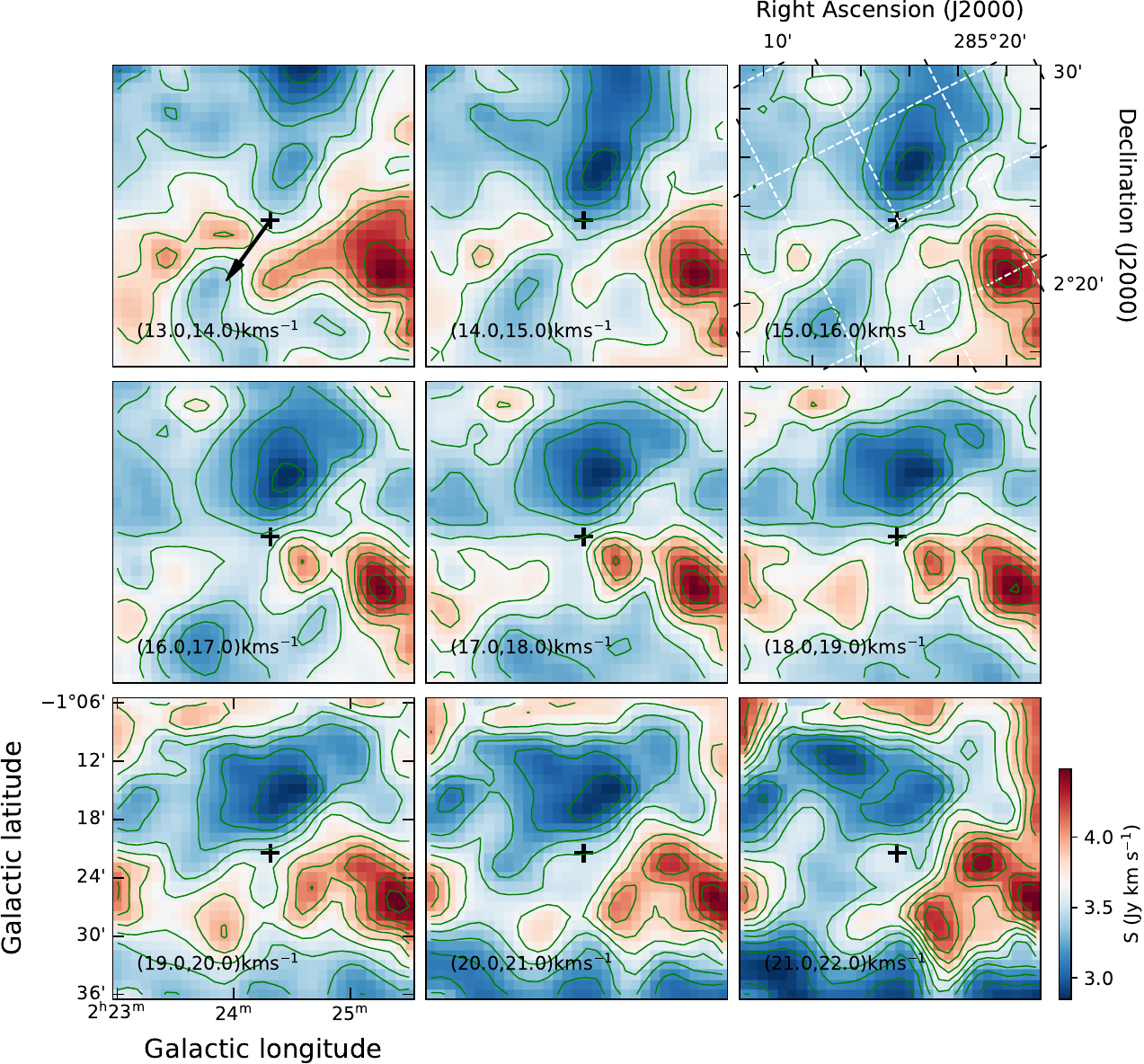}
\caption{Same as Figure~\ref{fig:vul 10cmap} but for
Sh 2-71.
 \label{fig:SH2-71cm}}
\end{figure}

Sh\,2-71 ($l=36^{\circ}.05, b=-1^{\circ}.37$) is an extensive, diffuse nebula with a max-diameter of 132.4${\arcsec}$, indicative of a highly evolved and ancient object \citep{Tylenda2003AA}. 
In the lower Galactic latitude regions, the dense ISM likely facilitated the development of a well-defined cavity through the AGB stellar wind-ISM interactions. 
The systematic velocity of Sh\,2-71 is 41.6\,km\,s$^{-1}$ \citep{Schneider1983ApJS}. 
Based on \cite{Fajin2022JPhCS}'s modeling, they further concluded that the expansion velocity
of the torus is V$_{exp}$\,=\,15\,km\,s$^{-1}$. 
Observing clear signs of toroidal structure expansion, we extracted the $\mathit{p\!-\!v}$ slice along the direction of proper motion, focusing on the torus expansion velocity range of $\sim$10~km~s$^{-1}$. 
This analysis, in Figure~\ref{fig:SH2-71m0}, revealed a potential large cavity opposite the direction of proper motion, which resembles an elongated, tail-like structure consistent with simulation results; this structure appears continuous within the velocity range from $V_{\rm LSR} = 13$~km~s$^{-1}$ to $V_{\rm LSR} = 17$~km~s$^{-1}$ (Figure~\ref{fig:SH2-71cm}).

A significant offset of $\sim 0\arcmin6$ is observed between the central star and the geometric center of the bubble, in the left panel of Figure~\ref{fig:SH2-71m0}. 
This can be attributed to the high \ion{H}{1} density along the upstream direction of the nebula's motion, which compresses the circumstellar material and distorts the bubble geometry. 
Assuming a constant expansion velocity of 15 km s$^{-1}$, this nebular dynamical age is calculated as $t_{\rm dyn} = \Theta_{\rm neb}\times d / V_{\rm exp}$, where $\Theta_{\rm neb}$ is the nebular diameter and $d$ is the distance, yielding $5.56\times10^{4}$ yrs. 
Combining $t_{\rm dyn}$ with the proper motion of 2.113 mas\,yr$^{-1}$ from \cite{EDR32021AA} predicts an angular displacement of $\sim1.96\arcmin$, while the $\!p\!-\!v$ diagram reveals an actual offset of $\sim3\arcmin39$. 
This discrepancy can be explained if the nebular proper motion has undergone deceleration. 
The bubble expansion timescale, calculated as $t_{\rm bubble} = \Theta_{\rm bubble} ({\rm arcsec}) \times d ({\rm pc}) / V_{\rm exp} ({\rm km\,s^{-1}})$ with $\Theta_{\rm bubble} = 3\arcmin39$, yields $\sim4.3\times10^{4}$ yr. 
This timescale aligns with $t_{\rm dyn}$ within the same order of magnitude, suggesting the bubble-carving wind originated during the proto-PN phase.

\subsection{Sh 2-68}

Sh\,2-68 exhibits a nebular diameter of $\sim\!400\arcsec$ from center to outer edge, consistent with an advanced evolutionary stage \citep{Acker1992ESO}. 
It is confirmed to have undergone PN–ISM interaction \citep{Mohery2022Ap&SS}, with a measured nebular expansion velocity of 6.6 km\,s$^{-1}$. The proper motion components are $-19.08\pm0.06$ mas yr$^{-1}$ in right ascension and $-56.34\pm0.05$ mas yr$^{-1}$ in declination \citep{EDR32021AA}, corresponding to tangential velocities $V_{\alpha}=60.679\pm0.19$ km s$^{-1}$ and $V_{\beta}=109.898\pm0.098$ km s$^{-1}$. 
This yields a total space velocity of 125.52 km s$^{-1}$, comparable to that of Sh\,2-188 as predicted by simulations \citep{wareing2006MNRAs}. 
Such high velocity enhances PN–ISM interaction and motivates searching for circumstellar bubbles formed by AGB wind–ISM interactions opposite the motion vector.

\begin{figure}[ht!]
\plotone{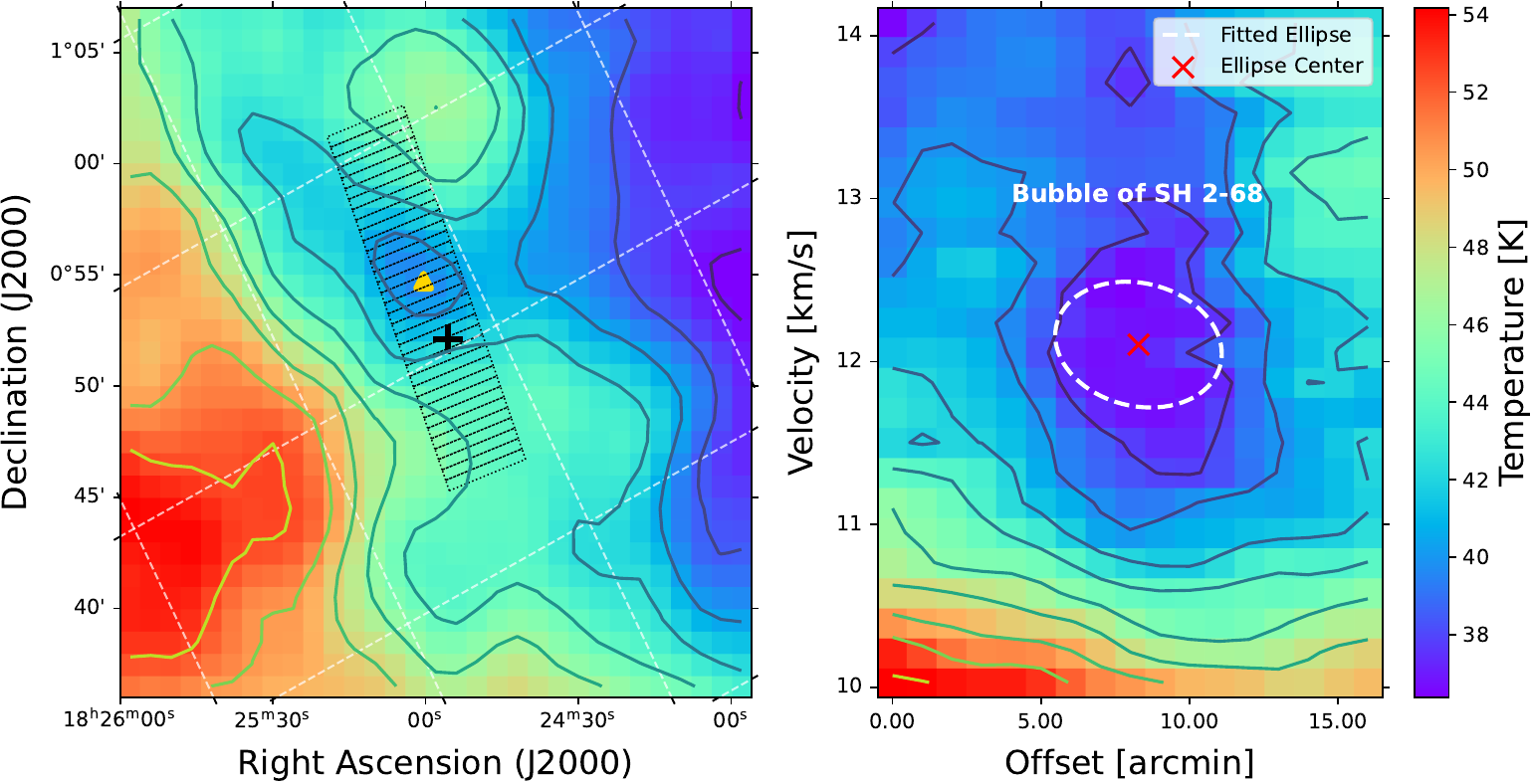}
\caption{Same as Figure~\ref{fig:K3-35m0} but for
Sh 2-68.
\label{fig:SH2-68m0}}
\end{figure}

\begin{figure}[ht!]
\plotone{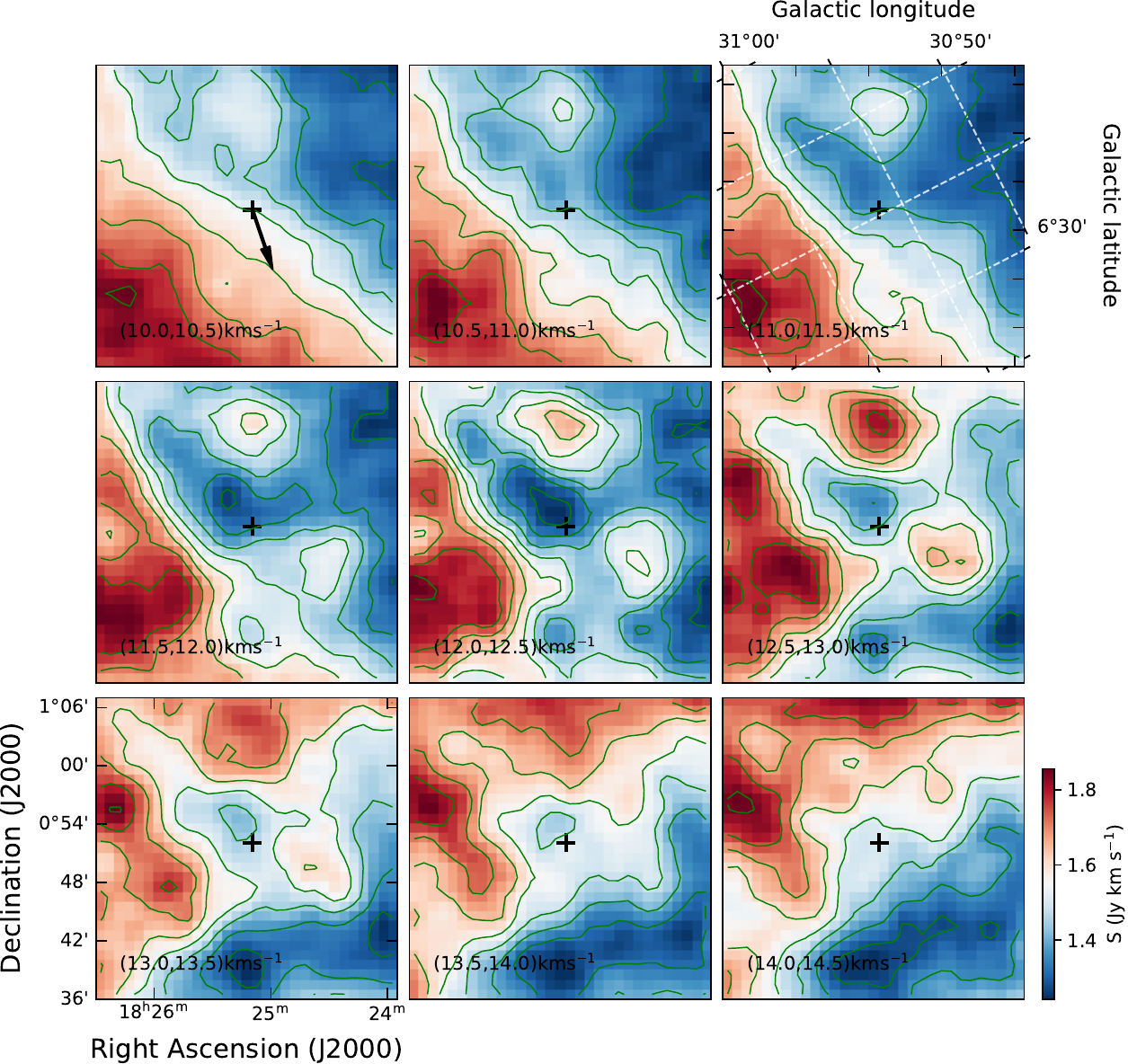}
\caption{Same as Figure~\ref{fig:vul 10cmap} but for
Sh 2-68.
\label{fig:SH2-68cm}}
\end{figure}

As shown in Figure~\ref{fig:SH2-68m0}, we identify a bubble downstream of the proper-motion direction within 9–14 km s$^{-1}$. 
The corresponding $p\!-\!v$ diagram reveals a conical morphology at 11--13 km s$^{-1}$, 
which is reminiscent of hydrodynamic structures reported in previous studies \citep[e.g.,][]{Arce2011ApJ}.
The channel maps (Figure~\ref{fig:SH2-68cm}) indicate that the bubble remains consistent across different velocities, presenting a cometary head-tail structure with asymmetric tail elongation in the velocity interval of 10.5--13.5 km s$^{-1}$. 
Collectively, these features provide robust evidence for the formation of a cavity resulting from the wind–ISM interaction.

\subsection{Abell 21}

\begin{figure}[ht!]
\plotone{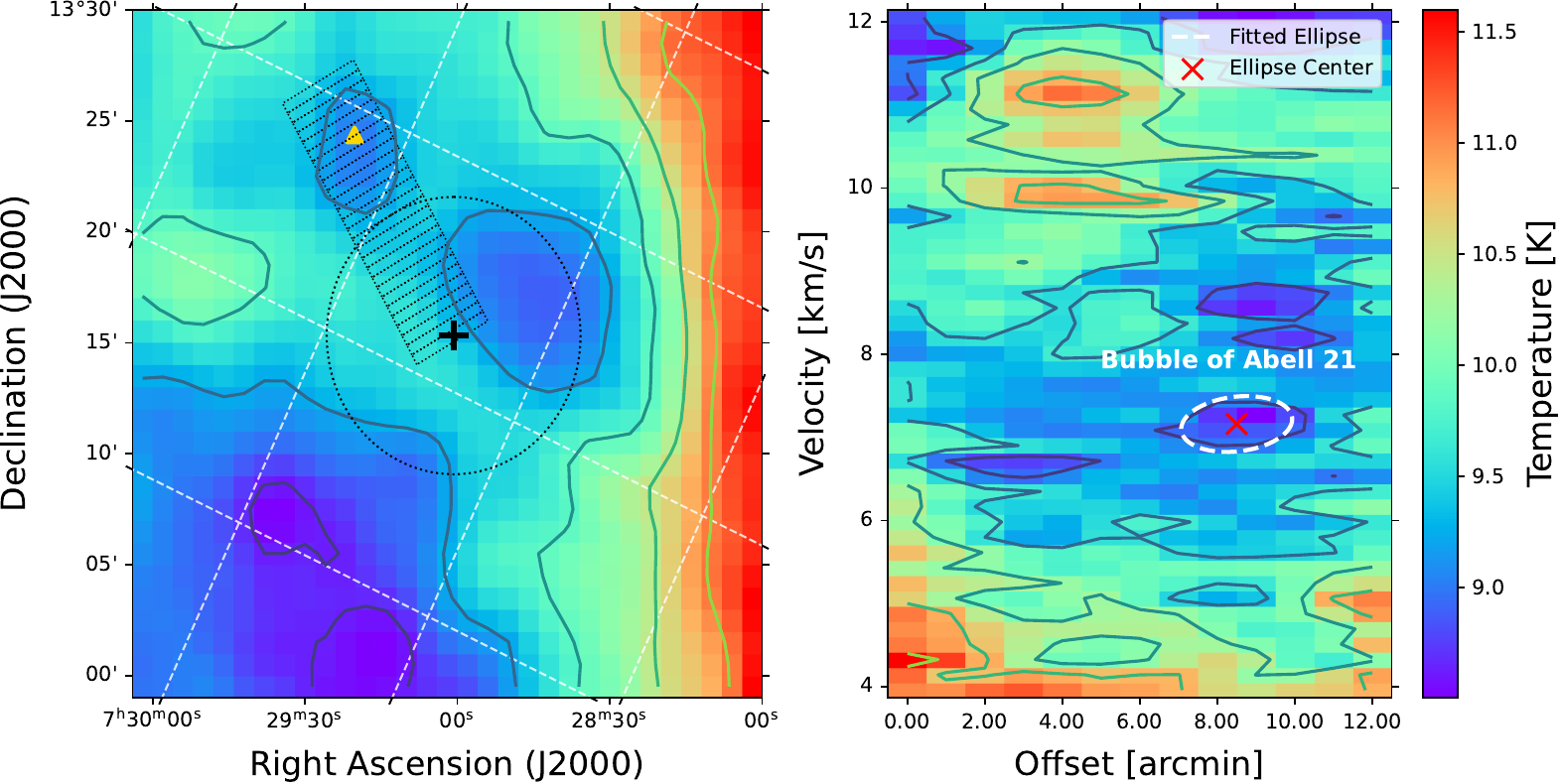}
\caption{Same as Figure~\ref{fig:K3-35m0} but for Abell 21. 
The black dotted circle marks the optical angular extent of Abell\,21, reported by \citetalias{FrewDavid2016MNRAS}. The white dashed grid displays the Galactic coordinate system. 
\label{fig:Abell21m0}}
\end{figure}

Abell\,21 lies at a moderate Galactic latitude. 
Previous trigonometric parallax measurements of Abell\,21 have yielded  a distance of 
$0.541^{+0.205}_{-0.117}$\,kpc \citep{Harris2007AJ}. 
Across the left and right panels of Figure~\ref{fig:Abell21m0}, we observe a bubble downstream of the nebula’s proper-motion direction, which exhibits an unusual elliptical morphology. 
This structure diverges from the typical geometries predicted by simulations of wind–ISM interactions \citep{Wareing2007MNRAS}, suggesting that additional physical mechanisms, such as magnetic fields or the presence of binary companions, may govern its dynamics \citep[e.g.,][]{sokernoam2001ApJ}. 
In the Milky Way, the scale height of atomic hydrogen (\ion{H}{1}) density decreases exponentially beyond $\sim$100\,pc from the Galactic disk \citep{Kalberla2009ARAA}. 
Consequently, PN-ISM interactions become less prominent at higher Galactic latitudes ($|b| >10^\circ$).  
This limits detectability of extended features, evidenced by the absence of fragmented substructures typically seen in denser ISM environments. 
Based on this analysis, we tentatively adopt the bottom bubble associated with Abell\,21.

\section{Discussion} \label{sec:apply}
\subsection{Assessment of the reliability of GRMs}

For a given PN located at a certain distance from the Sun, in the direction specified by Galactic coordinates ($l$, $b$), the relationship between the heliocentric distance $d$ and the Galactocentric distance $R$ can be derived from the following equation:  
\begin{equation}
    R^2 = R_0^2 + d^2 \cos^2 b - 2 R_0 d \cos b \sin l,       
\end{equation}  
where $R_0$ denotes the Sun's Galactocentric distance. 
Assuming circular orbits, the rotational velocity $\Theta(R)$ at a given Galactocentric distance $R$ is given by:  
\begin{equation}
    \Theta(R) = \frac{R}{R_0} \left( \frac{V_{\text{LSR}}}{\sin l \cos b} + \Theta_0 \right), 
\end{equation}  
where $V_{\text{LSR}}$ is the radial velocity relative to the LSR, and $\Theta_0$ is the rotational velocity of the Sun around the Galactic Center.  
The traditional Galactic rotational velocity profile is described by a power-law function, expressed as:  
\begin{equation}
    \Theta(R) = \Theta_0 \left[ a_1 \left( \frac{R}{R_0} \right)^{a_2} + a_3 \right]
\end{equation}  
where $a_1 = \alpha_1 \Theta_0$, and the parameters $a_2$ and $a_3$ are given in \citet{BrandGRM1993AA}.  
A ``Poly'' GRM is characterized by a fitted second-order polynomial, defined as follows:  
\begin{equation}
    \Theta(R) = a_1 + a_2 \rho + a_3 \rho^2, \quad
    R \in [4, 13]~\text{kpc}
\end{equation}
where $\rho = \left( \frac{R}{R_0} \right) - 1$, and the parameters $a_1$, $a_2$, and $a_3$ are provided in \citet{2014GRMReid2014ARAA}.  
Building on the power-law and polynomial GRMs introduced earlier, a simpler linear approximation of the Galactic Rotation Model was reported by \citet{Reild2009ApJGRM}, derived from fitted Galactic parameters and expressed as:  
\begin{equation}
    \Theta(R) = \Theta_0 + \frac{d\Theta}{dR}(R-R_0)
\end{equation}  
where the radial gradient   $\frac{d\Theta}{dR}$ is constrained to $2.3 \pm 0.9\,\text{km}\,\text{s}^{-1}\text{kpc}^{-1}$ over the Galactocentric distance range $R = 4$--$13\,\text{kpc}$.

A more comprehensive framework, the ``Univ'' universal rotation curve, was proposed by \citet{PersicGRM1996MNRAS} to incorporate contributions from both the exponential stellar disc and the dark matter halo. This model includes three adjustable parameters: $a_1$ (the circular rotation velocity at $R = 0.83R_{\text{opt}}$, where $R_{\text{opt}}$ denotes the optical radius of the Galaxy), $a_2 = R_{\text{opt}}/R$, and $a_3$ (a core-radius parameter characterizing the halo component). By fitting these three parameters, \citet{2014GRMReid2014ARAA} derived the following relation between $\Theta(R)$ and $R$:  
\begin{equation} \label{4}
    \Theta(R) = a_1 \cdot \left[
    \frac{1.97\,\beta\,x^{1.22}}{(x^2+0.78^2)^{1.43}} + (1-\beta)\,x^2\,\frac{1+a_3^2}{x^2+a_3^2}
    \right]^{1/2}
\end{equation}  
where $x = R/(a_2\,R_0)$ and $\beta = 0.72 + 0.44\,\log\left[(a_3/1.5)^5\right]$.  
Subsequently, \citet{ReildGRM2019ApJ} updated Equation (\ref{4}) to improve its accuracy, leveraging 200 trigonometric parallaxes and proper motions of molecular masers. The revised formula is given by:  
\begin{equation}
\Theta(R) = 200\,\lambda^{0.41} \cdot \left[
    \frac{
        \left(0.72 + 0.44\,\log_{10}\lambda\right) \cdot \frac{1.97\,x^{1.22}}{\left(x^2 + 0.61\right)^{1.43}} 
        + 1.6\,e^{-0.4\lambda} \cdot \frac{x^2}{x^2 + 2.25\,\lambda^{0.4}}
    }{
        0.8 + 0.49\,\log_{10}\lambda + \frac{0.75\,e^{-0.4\lambda}}{0.47 + 2.25\,\lambda^{0.4}}
    }
\right]^{1/2}
\end{equation}  
where $\lambda=(a_3/1.5)^5$ and $x = R/(a_2\,R_0)$ (consistent with the definition in Equation \ref{4}). Details of the different Galactic parameters corresponding to these models are listed in Table \ref{tab:GRMparameters}.

\begin{figure}[ht!]
\plotone{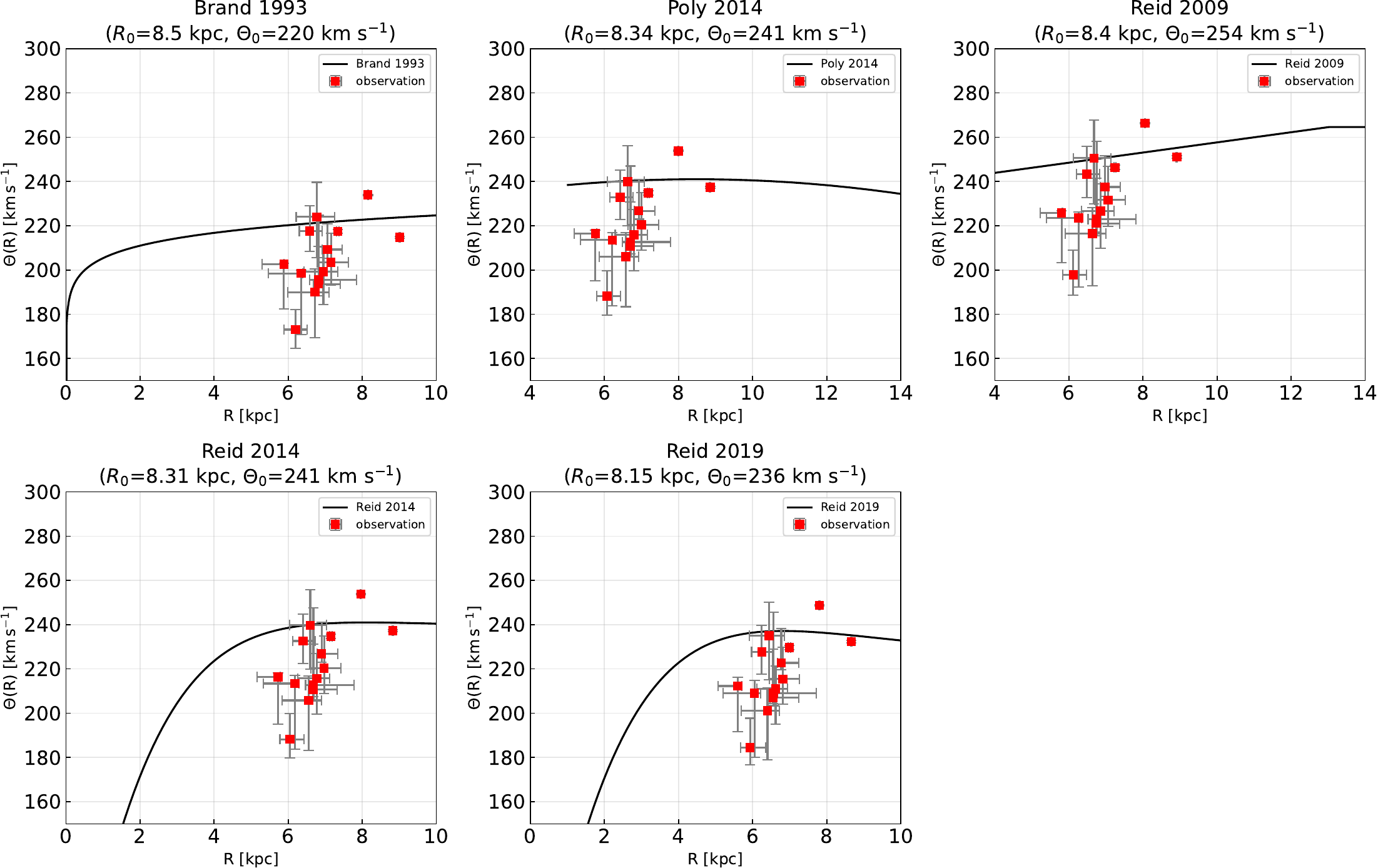}
\caption{Rotational velocity $\rm \Theta(R)$ versus Galactocentric radius R across different GRMs.
The  squares denote the observed values obtained based on
the calibrated parallax distances from
\citetalias{caGaiaEDR32021AJ}  and the bubbles' $\rm V_{lsr}$ under
different GRMs.
\label{fig:5GRM_rotation_R}}
\end{figure}

Figure~\ref{fig:5GRM_rotation_R} shows the different GRMs, each depicting the relationship between rotational velocity and Galactocentric radius. 
To evaluate the distinct GRMs, we adopt two sets of observational constraints: the central velocities of bubbles associated with PNe and the distances from \citetalias{caGaiaEDR32021AJ}. To quantify each model’s performance, we compare its predictions against these observational constraints and use the following metric:
\begin{equation}
Q = \sqrt{N^{-1} \sum^{N}_{i=1} \left[ \left( \frac{\Delta v_{i}}{\sigma_{v,i}} \right)^2 + \left( \frac{\Delta d_{i}}{\sigma_{d,i}} \right)^2 \right]},
\end{equation}
where $N$ denotes the number of sources; $\Delta v_{i}$ (in km s$^{-1}$) and $\Delta d_{i}$ (in kpc) are the deviations in velocity and distance, respectively, between the predictions of each GRM and the observational data for the $i$-th source; and $\sigma_{v,i}$ and $\sigma_{d,i}$ represent the observational uncertainties in velocity and distance for the $i$-th source.
Treating $Q$ as a metric of model goodness-of-fit, we find that the GRMs reported by \citet{BrandGRM1993AA} and \citet{ReildGRM2019ApJ} exhibit  relatively lower
fitting accuracy (see Table~\ref{tab:GRMparameters}). In contrast, the GRM presented in \citet{2014GRMReid2014ARAA} yields the best performance when tested against our bubble observational data.

\begin{figure}[ht!]
\plotone{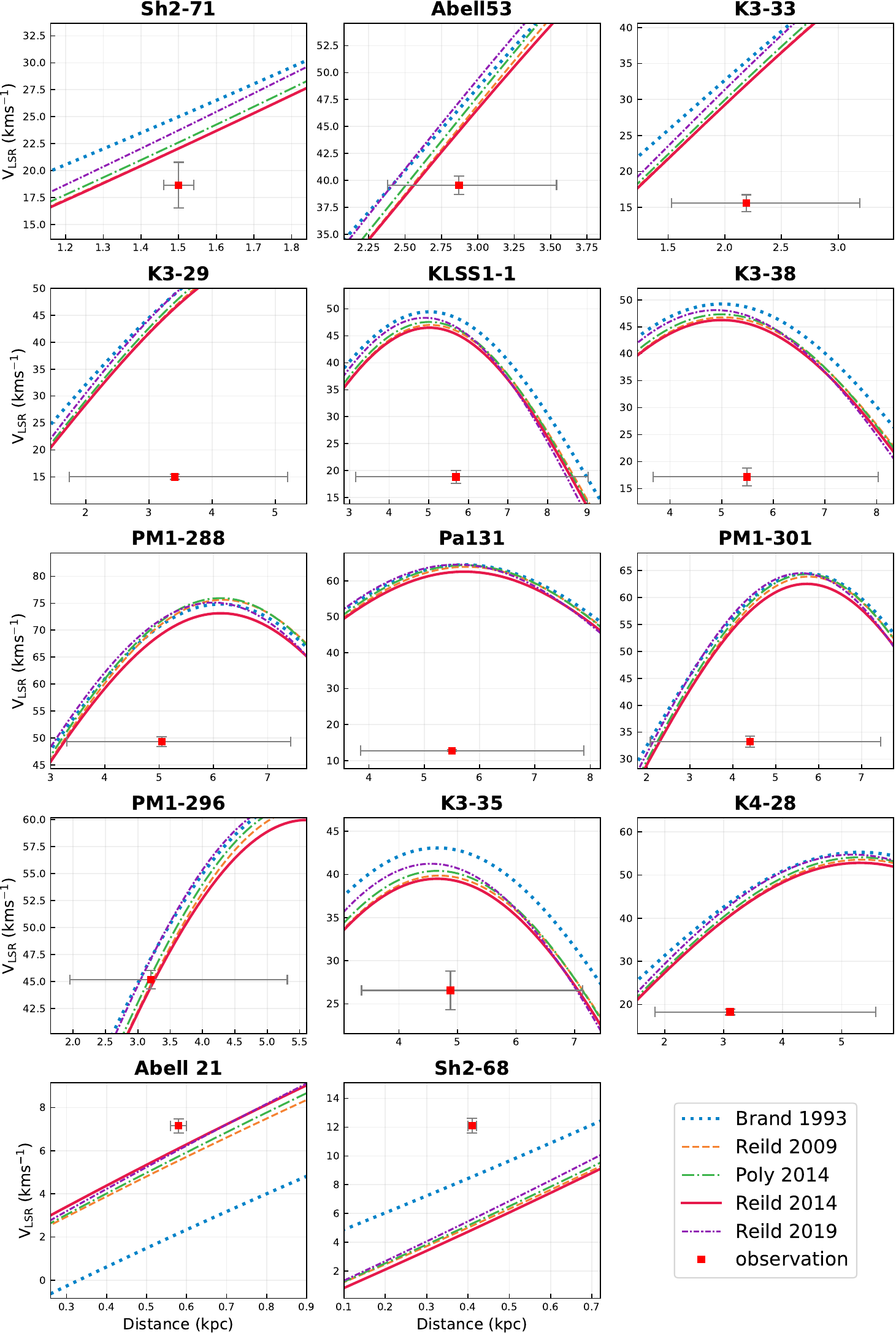}
\caption{ $\rm V_{lsr}$ of bubbles  against their distances derived from \citetalias{caGaiaEDR32021AJ}, with data shown for different directions. The curves illustrate the relational trends predicted by different GRMs, while the square symbols represent the observed values.
\label{fig:5GRM_Vlsr_dis}}
\end{figure}

Figure~\ref{fig:5GRM_Vlsr_dis} presents the LSR velocities of bubbles plotted as a function of distances to PNe, with data organized by distinct Galactic coordinate.  Overall, the observational measurements of the bubbles exhibit broad consistency with the predictions of the GRMs, though a subset of sources displays significant discrepancies.
Several factors may contribute to these observed deviations. One potential explanation lies in the kinematic complexity of the local ISM \ion{H}{1}. The GRMs primarily describe large-scale, axisymmetric Galactic rotation, but the local ISM can exhibit substructures (e.g., turbulent motions, non-circular flows, or localized density perturbations) that diverge from this global framework. Such small-scale kinematic deviations could introduce mismatches between the observed \ion{H}{1} velocities of bubbles and the large-scale rotational predictions of GRMs.
A second contributing factor may stem from uncertainties in PN distance estimates. For some PNe in the sample, the central stars that are critical for distance determination via methods like parallax or photometric scaling are obscured by extended nebular material, which can corrupt astrometric or photometric measurements and introduce errors into distance calculations. 
Additionally, the central stars of some PNe in our sample may be misidentified;
this ambiguity could arise from challenges in resolving central sources amid extended nebular emission or confusion with foreground/background stars. Furthermore, we cannot rule out the scenario that some bubbles have been misassociated with their respective PNe, as projection effects in the Galactic plane or incomplete kinematic linking between bubbles and PNe may lead to spurious pairings.

\subsection{Kinematic distance of PNe}

ISM bubbles physically associated with PNe share a coeval origin and are thus expected to trace the same Galactic kinematic environment as their host PNe. This allows the bubble’s LSR velocity to serve as a kinematic proxy for the PN. GRMs provide the key link between this velocity measurement and distance. Specifically, once the bubble’s LSR velocity is determined, it is fed into a well-constrained GRM to solve for the bubble’s radial distance from the Galactic center, thereby obtaining the distance of the associated PNe. In this study, we adopt the GRM refined via Method C proposed by \citet{kinematicdistancewenger2018ApJ}, which  is based on the Galactic rotation parameters presented in \cite{2014GRMReid2014ARAA}.

The uncertainties of the distance estimation arise primarily from two sources. First, velocity uncertainties originate from the ellipse fitting of the major and minor axes in the $p\!-\!v$ diagram, where the $\sim5$ km s$^{-1}$ velocity spreads observed in these diagrams (see Figures \ref{fig:K3-35m0} and \ref{fig:SH2-71m0}) primarily stem from the acceleration of the ISM by nebular expansion; this introduces significant uncertainty in determining the bubble's central velocity. 
Second, systematic uncertainties derive from the GRM, which carries inherent errors \citep[$\sigma_d/d \approx 25.8\%$ at the 1-$\sigma$ confidence interval; see Method C in][]{kinematicdistancewenger2018ApJ}. The average uncertainty of our kinematic distance (31\%) is consistent with that reported by \citetalias{FrewDavid2016MNRAS} (33\%). Notably, the systematic uncertainty of kinematic distances increases with larger distances due to error propagation in the GRM; small proper motion errors are amplified with increasing distance, and the limited accuracy of the rotation curve becomes more prominent at larger galactocentric radii.

\begin{figure}[ht!]
\plotone{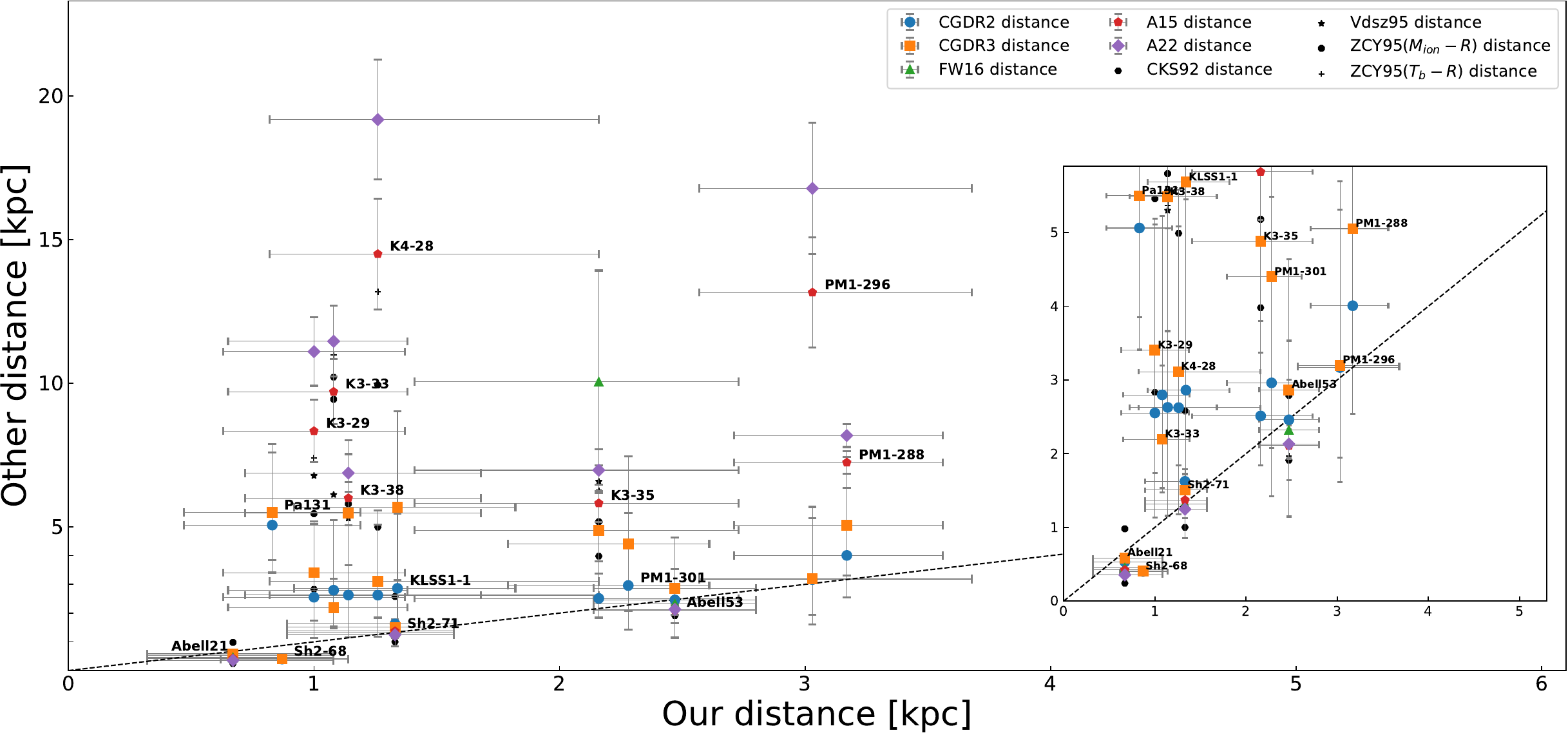}
\caption{Comparison between our measured distances with
others. The straight line denotes \(x = y\).
The inset is a zoom-in view. The strength of correlations with our distances, ranked in descending order, is: 
\citetalias{FrewDavid2016MNRAS} ($r=0.56$, $p=0.44$; 4 PNe); 
\citetalias{caGaiaDR22018AJ} ($r=0.37$, $p=0.19$; 14 PNe); 
\citetalias{caGaiaEDR32021AJ} ($r=0.32$, $p=0.27$; 14 PNe); 
\citetalias{Ali2015ApSS} ($r=0.20$, $p=0.58$; 10 PNe); 
\citetalias{AliA2022RAA} ($r=0.17$, $p=0.63$; 10 PNe); 
\citetalias{vdsz1995AA} ($r=0.09$, $p=0.88$; 5 PNe); \citetalias{CahnKalerStanghellini1992AAS} ($r=-0.10$, $p=0.81$; 8 PNe); \citetalias{ZhangCY1995ApJ} (M$_{\rm ion}$--$r$)($r=-0.12$, $p=0.78$; 8 PNe); and \citetalias{ZhangCY1995ApJ}(T$_{b}$--$r$) ($r=-0.14$, $p=0.74$; 8 PNe). 
\label{fig:comparingdistances}}
\end{figure}

As shown in Figure~\ref{fig:comparingdistances}, our derived distances are compared with  \citetalias{caGaiaDR22018AJ} and \citetalias{caGaiaEDR32021AJ} distances, alongside statistical distances based on different assumptions. 
Correlations are quantified using the Pearson coefficient ($r$) and null-hypothesis probability ($p$), with results for each method plotted in Figure~\ref{fig:comparingdistances}. 
Modest positive correlations ($r \geq 0.3$) are found with \citetalias{caGaiaDR22018AJ}, \citetalias{caGaiaEDR32021AJ}, and \citetalias{FrewDavid2016MNRAS} distances, while weak correlations appear for \citetalias{vdsz1995AA}, \citetalias{Ali2015ApSS}, and \citetalias{AliA2022RAA} distances. 
However, \citetalias{CahnKalerStanghellini1992AAS} and \citetalias{ZhangCY1995ApJ} distances show poor correlation. 
These differences likely reflect small sample sizes and systematic uncertainties in statistical distance methods.

Although \citetalias{caGaiaEDR32021AJ} distance uncertainties are reduced by $\sim$30\% compared to those of \citet{EDR32021AA}, they still suffer from large uncertainties for faint or highly extincted central stars. 
For example, \citetalias{caGaiaEDR32021AJ} distance uncertainty for PM\,1-296 is comparatively high ($\sim$60\%) due to its large apparent magnitude. 
In contrast, our derived kinematic distance not only agrees well with \citetalias{caGaiaDR22018AJ} and \citetalias{caGaiaEDR32021AJ}, but also provides a substantially smaller uncertainty. 
We thus demonstrate that estimated kinematic distances can provide a valuable alternative for PNe with large Gaia distance errors, offering an independent consistency check in cases of significant uncertainty.

The identification of the central star of Sh 2-71 remains controversial \citep{David2019MNRAS}. The nebular center, labeled A in \citet{David2019MNRAS}, is likely a close binary system that was designated as the central star of this PN by \citet{Kohoutek1979IBVS}. However, \citet{Frew2007apn4} proposed that the faint blue star northwest of A, labeled B in \citet{David2019MNRAS}, might be a more plausible candidate. The angular separation between A and B is approximately $\sim$7\arcsec.01. 
According to \citet{EDR32021AA}, the distances to stars at positions A and B are $1.60^{+0.05}_{-0.04}$ and $1.20^{+6.60}_{-0.56}$\,kpc, respectively. 
Our kinematic distance is closer to that of star B. 
This example demonstrates that when central star identification is contentious, measurements of 
\ion{H}{1} bubbles may provide an independent method to assess the reliable distance to the PNe.


\section{Conclusion} \label{sec:concl}

Using archival \ion{H}{1} data from the GALFA and GPPS surveys, we identified wind-carved bubbles associated with PNe. For the 14 bubble candidates we analyzed, we employed their kinematic properties to evaluate distinct GRMs and subsequently derived the distances to these PNe. Furthermore, a comparison of the GRMs indicates that the \citet{2014GRMReid2014ARAA} model performs better in fitting the data than that of \citet{BrandGRM1993AA}. The derived PN distances are in general agreement with previous measurements.
Our approach for estimating distances exhibits two key advantages: immunity to interstellar extinction and enhanced sensitivity to the prominent \ion{H}{1} 21-cm emission in the Galactic plane. While not inherently superior in accuracy to established methods, our approach provides an independent means of validating and resolving discrepant distance estimates for certain PNe. Notably, these kinematic distances are observationally limited to low Galactic latitudes, where the GRM reliably traces large-scale gas kinematics, and signatures of wind-ISM interactions are relatively prominent.

The identification of bubbles associated with PNe demands caution because foreground or background structures can easily form misleading projections that mimic genuine bubble-PNe associations. Future high-resolution observations will be essential to confirm the physical connection between the detected bubbles and their host PNe, while ruling out spurious alignments with these foreground or background structures.

\section*{Acknowledgements}

The financial supports of this work are from 
the National Natural Science Foundation of China (NSFC, No.\,12473027 and 12333005),
the Guangdong Basic and Applied Basic Research Funding (No.\,2024A1515010798), 
the Greater Bay Area Branch of the National Astronomical Data Center (No.\,2024B1212080003),
and the science research grants from the China Manned Space Project (NO. CMS-CSST-2021-A09, CMS-CSST-2021-A10, etc).
This work has used the data from the Five-hundred-meter Aperture Spherical radio Telescope (FAST). FAST is a Chinese national mega-science facility, operated by the National Astronomical Observatories of Chinese Academy of Sciences (NAOC).
This study utilized data from Galactic ALFA \ion{H}{1}  (GALFA \ion{H}{1} ) survey data obtained with the Arecibo 305 m telescope. 
The Arecibo Observatory is operated by SRI International under a cooperative agreement with the National Science Foundation (AST-1100968), and in alliance with Ana G. M{\'e}ndez-Universidad Metropolitana, and the Universities Space Research Association. The GALFA \ion{H}{1} surveys have been funded by the NSF through grants to Columbia University, the University of Wisconsin, and the University of California.





\bibliography{sample631}{}
\bibliographystyle{aasjournal}

\begin{deluxetable*}{ccccccccccccccc}
    \rotate
    \tabletypesize{\scriptsize}
    \renewcommand{\arraystretch}{1.6} 
    \caption{Positions, velocities, and distances of PNe.}\label{Tab: [table:alldistance]}
    \tablehead{
    \colhead{PN G} & \colhead{Name} & \colhead{$l$} & \colhead{$b$} & \colhead{$V_c$} & \colhead{Class} &
    \multicolumn{9}{c}{Distance (kpc)} \\
    \cline{7-15}
    \colhead{} & \colhead{} & \colhead{($^{\circ}$)} & \colhead{($^{\circ}$)} & \colhead{(km s$^{-1}$)} & \colhead{} &
    \colhead{\citetalias{caGaiaDR22018AJ}} & \colhead{\citetalias{caGaiaEDR32021AJ}} & \colhead{\citetalias{CahnKalerStanghellini1992AAS}} & \colhead{\citetalias{vdsz1995AA}} & \colhead{\citetalias{ZhangCY1995ApJ}} &
    \colhead{\citetalias{Ali2015ApSS}} & \colhead{\citetalias{AliA2022RAA}} & \colhead{\citetalias{FrewDavid2016MNRAS}} & \colhead{Ours}
    }
    \startdata
    035.9$-$01.1 & Sh\,2-71    & 36.054 & -1.367 & $18.66\pm2.12$ & A & $1.62^{+0.10}_{-0.09}$ & $1.50^{+0.04}_{-0.04}$ & 1.00 & 1.59 & 1.36 & $1.37^{+0.10}_{-0.10}$ & $1.25^{+0.12}_{-0.12}$ & $1.32^{+0.47}_{-0.47}$ & $1.33^{+0.24}_{-0.44}$ \\
    040.3$-$00.4 & Abell\,53  & 40.370 & -0.475 & $39.57\pm0.86$ & B & $2.49^{+2.18}_{-1.31}$ & $2.87^{+0.67}_{-0.49}$ & 1.91 & \nodata & 1.97 & $2.11^{+0.16}_{-0.16}$ & $2.32^{+0.20}_{-0.20}$ & $2.32^{+0.68}_{-0.68}$ & $2.47^{+0.33}_{-0.33}$ \\
    045.9$-$01.9 & K\,3-33   & 45.972 & -1.913 & $15.59\pm1.16$ & C & $2.80^{+2.43}_{-1.32}$ & $2.19^{+1.00}_{-0.66}$ & 10.12 & 6.12 & 10.98 & $9.70^{+1.13}_{-1.13}$ & $11.46^{+1.23}_{-1.23}$ & \nodata & $1.08^{+0.30}_{-0.43}$ \\
    048.1$+$01.1     & K\,3-29   & 48.162 &  1.163 & $15.01\pm0.46$ & B & $2.55^{+2.55}_{-1.42}$ & $3.41^{+1.79}_{-1.67}$ & 2.83 & 6.78 & 7.40 & $8.34^{+1.09}_{-1.09}$ & $11.10^{+1.19}_{-1.19}$ & \nodata & $1.00^{+0.37}_{-0.37}$ \\
    053.0+01.7     & KLSS\,1-1 & 53.094 &  1.765 & $18.84\pm1.22$ & C & $2.86^{+2.59}_{-1.55}$ & $5.69^{+3.34}_{-2.53}$ & \nodata & \nodata & \nodata & \nodata & \nodata & \nodata & $1.34^{+0.48}_{-0.42}$ \\
    053.2$-$01.5 & K\,3-38   & 53.200 & -1.534 & $17.16\pm1.61$ & B & $2.63^{+2.43}_{-1.47}$ & $5.49^{+2.54}_{-1.82}$ & 5.77 & 5.30 & 5.37 & $6.00^{+0.56}_{-0.56}$ & $6.88^{+0.66}_{-0.66}$ & \nodata & $1.14^{+0.54}_{-0.42}$ \\
    042.6$-$00.8 & PM\,1-288 & 42.663 & -0.865 & $49.32\pm0.93$ & A & $4.01^{+2.34}_{-1.46}$ & $5.05^{+2.37}_{-1.75}$ & \nodata & \nodata & \nodata & $7.24^{+0.40}_{-0.40}$ & $8.18^{+0.40}_{-0.40}$ & \nodata & $3.17^{+0.39}_{-0.46}$ \\
    046.6$-$01.4 & Pa\,131   & 46.688 & -1.436 & $12.74\pm0.22$ & A & $5.06^{+2.53}_{-1.65}$ & $5.50^{+2.38}_{-1.65}$ & \nodata & \nodata & \nodata & \nodata & \nodata & \nodata & $0.83^{+0.36}_{-0.36}$ \\
    046.7$-$01.6 & PM\,1-301 & 46.703 & -1.633 & $33.24\pm1.02$ & B & $2.96^{+2.53}_{-1.54}$ & $4.40^{+3.05}_{-2.32}$ & \nodata & \nodata & \nodata & \nodata & \nodata & \nodata & $2.28^{+0.33}_{-0.49}$ \\
    047.6$-$00.3 & PM\,1-296 & 47.688 & -0.302 & $45.17\pm0.86$ & A & $3.17^{+2.53}_{-1.55}$ & $3.20^{+2.11}_{-1.25}$ & \nodata & \nodata & \nodata & $13.16^{+1.91}_{-1.91}$ & $16.78^{+2.28}_{-2.28}$ & \nodata & $3.03^{+0.65}_{-0.46}$ \\
    056.0$+$02.0     & K\,3-35   & 56.096 &  2.094 & $26.57\pm2.21$ & A & $2.51^{+1.29}_{-0.67}$ & $4.88^{+2.26}_{-1.52}$ & 3.98 & 6.58 & 6.27 & $5.82^{+0.65}_{-0.65}$ & $6.97^{+0.72}_{-0.72}$ & $10.05^{+3.87}_{-3.87}$ & $2.16^{+0.57}_{-0.75}$ \\
    050.4$-$01.6 & K\,4-28   & 50.490 & -1.647 & $18.29\pm0.69$ & C & $2.62^{+2.46}_{-1.45}$ & $3.11^{+2.47}_{-1.27}$ & 4.99 & \nodata & 13.18 & $14.50^{+1.92}_{-1.92}$ & $19.17^{+2.08}_{-2.08}$ & \nodata & $1.26^{+0.39}_{-0.44}$ \\
    205.1$+$14.2     & Abell\,21 & 205.139 & 14.241 & $7.16\pm0.33$  & B & $0.53^{+0.02}_{-0.02}$ & $0.58^{+0.02}_{-0.02}$ & 0.24 & \nodata & 0.44 & $0.42^{+0.04}_{-0.04}$ & $0.35^{+0.04}_{-0.04}$ & $0.45^{+0.13}_{-0.13}$ & $0.67^{+0.41}_{-0.35}$ \\
    030.6$+$06.2     & Sh\,2-68   & 30.673 &  6.279 & $12.10\pm0.51$ & A & $0.40^{+0.01}_{-0.01}$ & $0.41^{+0.01}_{-0.01}$ & \nodata & \nodata & \nodata & \nodata & \nodata & \nodata & $0.87^{+0.27}_{-0.25}$ \\
    \enddata
    \tablecomments{$V_{c}$ represents the central velocity of the identified bubble.}
\end{deluxetable*}

\begin{deluxetable*}{cccccc}
    \tabletypesize{\scriptsize}
    \renewcommand{\arraystretch}{1.6} 
    \caption{Parameters of GRMs} \label{tab:GRMparameters}
    \tablehead{\colhead{Parameters}
     & \colhead{\citet{BrandGRM1993AA}} & \colhead{\cite{Reild2009ApJGRM}}  &\colhead{Poly} & \colhead{\citet{2014GRMReid2014ARAA}} & \colhead{\citet{ReildGRM2019ApJ}}
    }
    \startdata
    $R_0$\,(kpc) & 8.5 & $8.4\pm 0.86$ & $8.34\pm0.17$ & $8.31\pm0.16$ & $8.15\pm0.15$ \\
    $\Theta_0$\,(km\,s$^{-1}$) & 220 & $254 \pm 16$ & $241\pm9$ & $241\pm8$ & $236\pm7$ \\
    $a_1$\,(km\,s$^{-1}$) & 221.69 & \nodata & $241\pm9$ & $241\pm8$ & \nodata \\
    $\alpha_1$ & 1.0077 & \nodata & 1 & 1 & \nodata \\
    $a_2$ & 0.0394 & \nodata & $0.5\pm3.7$ & $0.9\pm0.06$ & $0.96\pm0.05$ \\
    $a_3$ & 0.0071 & \nodata & $-15.1\pm8.4$ & $1.46\pm0.16$ & $1.62\pm0.02$ \\
    \hline
    $Q$ & 68.9 & 68.0 & 68.7 & 66.5 & 69.2 \\
    \enddata
\end{deluxetable*}

\newpage

\end{document}